\documentclass[preprint,aps,prd,showpacs,nofootinbib,superscriptaddress,tightenlines]{revtex4}
\usepackage{amsfonts}
\usepackage{mathrsfs}
\usepackage{graphicx}
\usepackage{amsmath}
\usepackage{amssymb}
\usepackage{bm}
\usepackage{bbm}
\usepackage{multirow}
\def\hlinew#1{%
  \noalign{\ifnum0=`}\fi\hrule \@height #1 \futurelet
   \reserved@a\@xhline}

\def\OMIT#1{}

\newcommand{\nn}{\nonumber}

\newcommand{\beq}{\begin{equation}}
\newcommand{\eeq}{\end{equation}}
\newcommand{\bqa}{\begin{eqnarray}}
\newcommand{\eqa}{\end{eqnarray}}

\begin{document}

\title{\mbox{}\\[10pt]
Exclusive decay of $P-$wave Bottomonium into double $J/\psi$}

\author{Juan Zhang\footnote{E-mail: xzj\_123789@163.com}
} \affiliation{Institute of High Energy Physics, Chinese Academy of
Sciences, Beijing 100049, China\vspace{0.2cm}}
\affiliation{Institute of Theoretical Physics, Shanxi University,
Taiyuan, Shanxi 030006, China\vspace{0.2cm}}

\author{Hairong Dong\footnote{E-mail: donghr@ihep.ac.cn}
}
\affiliation{Institute of High Energy Physics, Chinese Academy of
Sciences, Beijing 100049, China\vspace{0.2cm}}

\author{Feng Feng
\footnote{E-mail: fengf@ihep.ac.cn} }
 \affiliation{Theoretical
Physics
Center for Science Facilities, Institute of High Energy Physics,\\
Chinese Academy of Sciences, Beijing 100049, China\vspace{0.2cm}}
\affiliation{Institute of High Energy Physics, Chinese Academy of
Sciences, Beijing 100049, China\vspace{0.2cm}}

\date{\today}
\begin{abstract}

We calculate the relativistic corrections of $J/\psi$, including electromagnetic corrections, to $\chi_{bJ}\to J/\psi J/\psi$ in the framework of nonrelativistic QCD factorization. The relativistic effect is found to increase the lower-order prediction for the decay width by about $10\%$, while the electromagnetism contribution is very small, about $0.2\%$ for $\chi_{b0}$ and $\chi_{b2}$. The total branching ratio is predicted to be of order $10^{-5}$ for $\chi_{b0,b2}\to J/\psi J/\psi$, but
$10^{-11}$ for $\chi_{b1}\to J/\psi J/\psi$, since there is only electromagnetism contribution in this channel. We predict it is possible to observe these reactions in LHC. Finally, we  estimate the decay width and branching ratio of $\chi_{cJ}\to \omega\omega,\phi\phi$ in the constituent quark model by our formula at the leading-order of relativistic correction and electromagnetic correction. The obtained branch ratio of $\chi_{c0,2}\to \phi\phi$ is in agreement with the experimental measurement in the order of magnitude.
\end{abstract}

\pacs{\it  12.38.-t, 12.38.Bx, 13.20.Gd, 12.39.Pn}
\maketitle

\section{Introduction}
The nonrelativistic nature of heavy quarkonium provides people with a valid component to understand the nonrelativistic effect of QCD. Since the quarkonium includes several well-separated scales which contain both hard scale and soft scale, the studies of hadronic reactions involving heavy quarkonium are significative of giving us an insight into both perturbative and nonperturbative QCD.

The running of $B$ factories with high luminosity has made the measurement of hadronic-exclusive processes feasible and given the results that raised new challenges to the existing theory. One of the largest puzzles is the cross section for exclusive double charmonium-production in annihilation i.e. $e^+e^-\to J/\psi +\eta_c$, at the $B$ factory energy of $\sqrt{s}=10.6 \ \mathrm{GeV}$(Ref.~\cite{Abe:2002rb,Abe:2004ww,Aubert:2005tj}) is about an order of magnitude larger than the leading-order nonrelativistic QCD(NRQCD) predictions in Refs.~\cite{Braaten:2002fi,Liu:2002wq,Liu:2004ga}.

The NRQCD factorization~\cite{Bodwin:1994jh} as an outstanding effective field theory approach to dealing with the physical problem involving multiple scales provides us with a systematic framework to deal with the exclusive quarkonium production process; the amplitude can be factorized as the products of the short-distance, perturbatively calculable coefficients and the long-distance, nonperturbative NRQCD matrix elements which are universal for all processes. The short-distance coefficient can be expanded by the order in $\alpha_s$, while the NRQCD matrix elements are organized as a series in the relative velocity $v$ of the heavy quark, so one can improve the NRQCD predictions simultaneously in $\alpha_s$ and $v$. Considering the next-to-leading-order perturbative corrections to $e^+e^-\to J/\psi+\eta_c$, the discrepancy between theory and experiment measurement is greatly alleviated~\cite{Zhang:2005cha,Gong:2007db,Hagiwara:2003cw, He:2007te, Bodwin:2007ga, Braguta:2008tg}.

In recent years, the studies of double-charmonium production at $B$ factory have already obtained huge theoretical progress in understanding hadronic exclusive processes with charmonium production. These facts inspire us to explore more analogous and valuable processes to add to our knowledge about these exclusive processes involving quarkonium. Similar to $e^+e^-$ annihilation, double-charmonium production in bottomonium decays can also be used to study the dynamics of hard exclusive processes and the structure of charmonium mesons; the reason is that the bottomonium can decay into not only light hadron but also charmonium because its mass is heavier than charmonium and close to the energy of $B$ factory. Although the investigations about bottomonium decay are not as intense as double-charmonium production in annihilation, there are several works focusing on these decays~\cite{Braguta:2005gw, Hao:2006nf, Jia:2007hy, Braaten:2000cm,Maltoni:2004hv,Jia:2006rx,Gong:2008ue,Braguta:2009xu,Sun:2010qx}. In Ref.\cite{Braaten:2000cm,Maltoni:2004hv,Jia:2006rx,Gong:2008ue,Braguta:2009xu,Sun:2010qx}, the double-charmonium production in exclusive bottomonium decays $\eta_b \to J/\psi J/\psi$ has been investigated. These researches groped the potential of the discovery of this hadronic decay channel in experimentation. As the $S$-wave bottomonium decays into double $J/\psi$ have been researched, it is natural to carry on a further study of the similar $P$-wave bottomonium decays into double $J/\psi$.

 This paper is intended to deal with the $P$-wave bottomonium decay process $\chi_{bJ}\to J/\psi J/\psi$, which is a part of the processes dealt with in Refs.\cite{Braguta:2010zz,Braguta:2005gw}. In Ref.\cite{Braguta:2005gw}, light-cone method is employed, and only the QCD process is considered. However, using NRQCD factorization to handle this type of process with the initial and final states all involving heavy quarkonium is more natural. In this paper, we calculate the relativistic corrections to this process in the NRQCD frame; the pure QED process is also considered. In fact, this work is an extended and updated version of Ref.~\cite{zhangjuan}.

This paper is organized as follows. In Sec.~\ref{factorization}, we describe the NRQCD factorization formula relevant to this work and compare our matching scheme with the orthodox doctrine. We also present a detailed description on how to determine the short-distance coefficients through relative order $v^2$ in $\chi_{bJ}\to J/\psi J/\psi$. In Sec.~\ref{fullqcd}, we perform calculations of the amplitudes for $\chi_{bJ}\to J/\psi J/\psi$ that include the relativistic corrections to the order of $v^2$ of $J/\psi$, helicity amplitudes and electromagnetism contributions. In Sec.~\ref{numericalresult}, we apply our formulas to investigate the phenomenological impact of QCD and QED corrections to the decay width and branching ration of $\chi_{bJ}\to J/\psi J/\psi$. By analyzing the numerical results, We predict the possibility of the observation of $\chi_{bJ}\to J/\psi J/\psi$ decay in the experimentation. Finally, we summarize our results in Sec.~\ref{summarization}.

\section{NRQCD factorization and matching strategy}
\label{factorization}
Heavy quarkonium is a nonrelativistic system involving multiple scales. Because the velocity $v$ of the heavy quark is much less than $1$, the following scales are well separated: the heavy-quark mass $m_Q$, the relative momentum $m_Qv$, and the binding energy $m_Qv^2$. NRQCD factorization provides us with a valid effective field theory to separate the scale $m_Q$ from the others. In the NRQCD factorization formula, the contribution from the scale that is less than $m_Q$ is kept in the Lagrange operators, while the contribution from the scale that is more than $m_Q$ is absorbed in the Wilson coefficients. With NRQCD factorization, the decay width can be expressed as the sum over $q\bar q$ channels of products of a long-distance-physics-insensitive short-distance coefficient and a process independent long-distance non-perturbative
matrix element.

 For a $Q(p)\bar{Q}(\bar{p})$ pair with total momentum $P$ and relative momentum $q$ we express the momenta of $c$ and $\bar{c}$ in perturbative matching:
%
\bqa
p=\frac{P}{2}+q,\ \ \ \ \ \bar{p}=\frac{P}{2}-q,
\eqa
%
where $q$ and $P$ satisfy $P_i\cdot q_i=0$, and
$p^2=\bar{p}^2=m^2_Q$. $P$ is the true total momentum of the $c\bar c$ pair,
$P=p+\bar{p}$, with invariant mass of $2 E_q$. In the rest frame of the
$c\bar{c}$ pair, the explicit components of the momenta are
$P^\mu=(2E_q,\mathbf{0})$, $q^\mu=(0,\mathbf{q})$,
$p^\mu=(E_q,\mathbf{q})$, and $\bar{p}^\mu=(E_q,-\mathbf{q})$,
respectively.

 To be accurate to order $v^2$, there are two methods that can be used to match the short-distance coefficients. One is the traditional matching method, in which we need to expand the energy of the $Q$ or the $\bar Q$ in the $Q\bar Q$ rest frame $E_q$ around the pole mass $m_c$ in power series of $\mathbf{q}^2$,
\begin{equation}\label{eq}
E_q = m_c +
{{\mathbf q}^2\over 2 m_c}+{\mathcal O}({\mathbf q}^4).
\end{equation}
The other method which we used in this paper is to expand every occurrence of $m_c$ in the amplitude in terms of ${\mathbf q^2/E_q^2}$, while keeping $E_q$ intact:
\begin{equation}
m_c = E_q - {{\mathbf q}^2\over 2 E_q}+{\mathcal O}({\mathbf q}^4).
\end{equation}
In the first method, when summing the polarization states of $c\bar{c}({}^3S_1)$, $b\bar{b}({}^3P_J)$, there are new factors of $E_q$, which are regenerated in the squared amplitude, we have to reexpand these occurring $E_q$ factors again and realign the corresponding terms from the leading order (LO) matrix element squared to the relativistic correction piece. The second method avoids many complications that emerged in the first one and eliminates the task of matching the cross section to the amplitude squared.

There is another question that needs to be considered. In our matching method, $m_c$ has been eliminated in favor of $E_q$ in the physical matrix element squared, then we need decide which value of $E_q$ should be taken to give the prediction. After comparing the Eq. (\ref{eq}) which comes from simple nonrelativistic kinematics and the G-K relation~\cite{Gremm:1997dq}:
\bqa
{M_{J/\psi}\over 2 m_c} &=& 1+ {1\over 2} \langle v^2 \rangle_{J/\psi} + O(v^4),
\label{G-K:relation:first}%
\eqa
where $\langle v^2\rangle_{J/\psi}$ is a dimensionless ratio of the vacuum matrix elements defined as follow:
\bqa
\langle v^2 \rangle_{J/\psi}  \approx
{\langle J/\psi(\lambda)| \psi^\dagger
(-\tfrac{i}{2}\tensor{\mathbf{D}})^{2}
\bm{\sigma}\cdot\bm{\epsilon}(\lambda)\chi|0\rangle \over m_c^2\,
\langle J/\psi(\lambda)| \psi^\dagger
\bm{\sigma}\cdot\bm{\epsilon}(\lambda)\chi|0\rangle}.
\label{v2:jpsi:definition}%
\eqa
We can see that theoretical consistency requires that $E_q$ can be fixed in an unambiguous manner, i.e. $E_q$ appearing everywhere in the short-distance coefficients can be replace by $M_{J/\psi}/2$. By this way, the relativistic effects in phase space integrals are automatically incorporated. In addition, choosing $M_{J/\psi}$ as the input parameter is better than $m_c$ since the mass of $J/\psi$ is known rather precisely while the charm quark mass is ambiguously defined.

Since we no longer need to worry about the complication from the phase space integral and sum of the polarization states, we can match the short-distance coefficients at the amplitude level. Because they are similar for the $\chi_{b0,b1,b2}\to J/\psi J/\psi$, here we take the $\chi_{b0}$ as an example to demonstrate how to match the short-distance coefficients. At the leading order of $\alpha_s$ and the order of $v^2$, the amplitude $\mathcal {M}[\chi_{b0}\to J/\psi J/\psi]$ in terms of the vacuum-to-$J/\psi$ and $\chi_{b0}$-to-vacuum matrix elements can be written as:
\bqa\label{nrqcd-exp:CS:ampl}
\mathcal {M}_{\chi_{b0}}
&=&\sqrt{2M_{\chi_{b0}}}\sqrt{2M_{J/\psi_1}}\sqrt{2M_{J/\psi_2}}\sum_{m,n=0}^{m,n=1}
c_{mn}^0\langle
J/\psi_1|\psi^\dag\left(-\frac{i}{2}\tensor{\bm{D}}\right)^{2m}
\bm{\sigma}\cdot\bm{\epsilon}(\lambda_1)\chi|0\rangle\nn\\
&&\langle
J/\psi_2|\psi^\dag\left(-\frac{i}{2}\tensor{\bm{D}}\right)^{2n}
\bm{\sigma}\cdot\bm{\epsilon}(\lambda_2)\chi|0\rangle\frac{1}{\sqrt{3}}\langle0|\chi^\dag
(-\tfrac{i}{2} \tensor{\bm{D}}\cdot \bm{\sigma}\
)\psi|\chi_{b0}\rangle\,,
\eqa
where $c_{mn}^0$ are the corresponding short-distance coefficients, which are Lorentz scalars formed by various kinematic invariants in the reaction. In particular, they also depend explicitly on the helicity $\lambda$ of $J/\psi$. For the Lorentz-invariant amplitude in the left-hand side of Eq.~(\ref{nrqcd-exp:CS:ampl}), $\mathcal{M}_{\chi_{b0}}$, it is most natural to assume relativistic normalization for each particle state, since the
squared amplitude needs to be folded with the relativistic phase space integral to obtain the physical decay width. However, in the right-hand side of Eq.~(\ref{nrqcd-exp:CS:ampl}), the $J/\psi$ and $\chi_{bJ}$ state appearing in the NRQCD matrix elements conventionally assume the nonrelativistic normalization. To compensate this difference, one must insert a factor $\sqrt{2 M_{J/\psi}}\sqrt{2 M_{J/\psi}} \sqrt{2 M_{\chi_{bJ}}}$ in the right side of Eq. (\ref{nrqcd-exp:CS:ampl}).

To determine the coefficients $ c_{mn}^0$we follow the moral that these short-distance coefficients are insensitive to the long-distance confinement effects, so one can replace the physical $J/\psi$ state by a free $c\bar{c}$ pair of quantum number ${}^3S_1$, and replace the physical $\chi_{bJ}$ state by a free $b\bar{b}$ pair of quantum number ${}^3P_J$, by which the NRQCD operator matrix elements can be perturbatively calculated. The short-distance coefficients
$c_{mn}^J$ can then be read off by comparing the QCD amplitude for $\mathcal {M}[b\bar{b}({}^3P_J)\to c\bar{c}({}^3S_1,P_1,\lambda_1)+c\bar{c}({}^3S_1,P_2,\lambda_2)]$ and the corresponding NRQCD factorization formula.

After simple NRQCD calculation, the perturbative NRQCD matrix elements for $Q\bar Q({}^3S_1)$ and $Q\bar Q({}^3P_1)$ states at leading order are simply expressed as:
%
\begin{subequations}
\bqa
&&\langle Q\overline{Q}({}^3S_1)|\psi^\dag\bm{\epsilon}\cdot\bm{\sigma}\chi|0\rangle=\sqrt{2N_c}\,,\\&&
\langle Q\overline{Q}({}^3S_1)|\psi^\dag\left(-\frac{i}{2}\tensor{\bm{D}}\right)^{2}
\bm{\sigma}\cdot\bm{\epsilon}\chi|0\rangle=\sqrt{2N_c}\,\mathbf{q}^2\,,\\
&&\frac{1}{\sqrt{3}}\langle0|\chi^\dag(-\tfrac{i}{2}
\tensor{\bm{D}}\cdot \bm{\sigma}\
)\psi|Q\overline{Q}({}^3P_0)\rangle
=\sqrt{2N_c}\,|\mathbf{q}|\,,
\eqa\label{eq:gh}
\end{subequations}
%
 where $\sqrt{2N_c}$ comes from the spin and color factors of the NRQCD matrix elements and the state $|Q\overline{Q}({}^{2s+1}L_J)\rangle$ is nonrelativistically normalized.

Using the formula above, we can obtain the partonic level amplitude $\mathcal {M}[b\bar{b}({}^3P_0)\to c\bar{c}({}^3S_1,P_1,\lambda_1)+c\bar{c}({}^3S_1,P_2,\lambda_2)]$ expanded to the order $v^2$:
\begin{subequations}
\bqa
\mathcal {M}_{{}^3P_0}
&=&8E(q_1)E(q_2)E(Q)
\sum_{m=0}^1\sum_{n=0}^1c_{mn}^0\frac{1}{\sqrt{3}}\langle0|\chi^\dag
(-\tfrac{i}{2} \tensor{\bm{D}}\cdot \bm{\sigma}\
)\psi|b\bar{b}({}^3P_0)\rangle \\&&\langle
c\bar{c}({}^3S_1,\lambda_1)|\psi^\dag\left(-\frac{i}{2}\tensor{\bm{D}}\right)^{2m}
\bm{\sigma}\cdot\bm{\epsilon}(\lambda_1)\chi|0\rangle
 \langle
c\bar{c}({}^3S_1,\lambda_2)|\psi^\dag\left(-\frac{i}{2}\tensor{\bm{D}}\right)^{2n}
\bm{\sigma}\cdot\bm{\epsilon}(\lambda_2)\chi|0\rangle\nn\\
&\approx&(2N_c)^{3/2}8E(q_1)E(q_2)E(Q)\left[c_{00}^0+c_{10}^0\frac{\mathbf{q}_1^{2}}{m_c^2}+c_{01}^0\frac{\mathbf{q}_2^{2}}{m_c^2}+\cdots\right]
\eqa
\label{ampl:ccbar}
\end{subequations}
%
In Eq.~(\ref{ampl:ccbar}), we use relativistic normalization for the
$c\bar c$ and $b\bar b$ states in the computation of the QCD
amplitude and nonrelativistic normalization in the NRQCD matrix
elements. Consequently, a factor $2E_q$ appears in the second
expression of Eq.~(\ref{ampl:ccbar}).

 From Eq. (\ref{ampl:ccbar}), it is straightforward to extract
the short-distance coefficients $c_{mn}^J$:
%
\bqa
c_{mn}^J=\frac{m_c^{2(m+n)}}{m!n!}\frac{\partial^m}{\partial
\mathbf{q}_1^{2m}}\frac{\partial^n}{\partial
\mathbf{q}_2^{2n}}\left.\left[\frac{\mathcal {M}[b\bar b({}^3P_J)\to c\bar
c(^3S_1,P_1,\lambda_1)+c\bar c(^3S_1,P_2,\lambda_2)] }{(2N_c)^{3/2}8
E(q_1)E(q_2)E(Q)}\right]\right\vert_{\mathbf{q}_1^{2}=\mathbf{q}_2^{2}=0}\,\label{eq:ab}
\label{shortcoefficient} \eqa
We can derive the LO coefficient $c_{00}^J$ by putting ${\mathbf
q}\to 0$ in the amplitude and equating $E_q$ and $m_c$. While
deducing the coefficient $c_{01}^J,c_{10}^J$, we need first expand
the amplitude to the first order in ${\mathbf q}^2$ prior to taking
the ${\mathbf q}\to 0$ limit.

\section{Color-singlet model calculation}\label{fullqcd}
In this section, we present a calculation for $\chi_{bJ}\to J/\psi J/\psi$ in perturbative QCD scheme. As we discussed in Sec.~\ref{factorization}, the short-distance coefficients are insensitive to the long-distance confinement effects, and to obtain the short-distance coefficients $c_{00}^J$ and $c_{01}^{J}$, $ c_{10}^{J}$, we need only to compare the QCD amplitude and the corresponding NRQCD factorization formula in $Q\bar Q$ state. So we replace the physical
$J/\psi$, $\chi_{bJ}$ states by a free $c\bar c$ pair of quantum number $^3S_1$ and a $b\bar b$ pair of quantum number $^3P_J$, respectively, and we compute the $Q\bar Q$ analog amplitude $\mathcal {M}_{^3P_J}$ of the hadronic amplitude $\mathcal {M}_{\chi_{bJ}}$, where the amplitude for $Q\bar Q$ level perturbative process $\mathcal {M}[b\bar{b}({}^3P_J)\to c\bar{c}({}^3S_1,P_1)+c\bar{c}({}^3S_1,P_2)]$ has been aliased as $\mathcal {M}_{^3P_J}$ and the hadronic level amplitude $\mathcal{M}[\chi_{bJ}\to J/\psi+J/\psi]$ has been aliased as $\mathcal {M}_{\chi_{bJ}}$.
\begin{figure}[h]
\begin{center}
\includegraphics[scale=0.8]{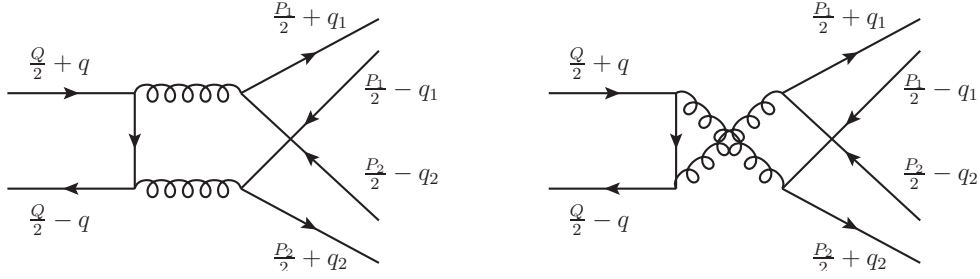}
\caption{ Leading-order QCD Feynman diagrams that contribute to
$\chi_{bJ} \to J/\psi J/\psi$ \label{feynman:diag:gamma}}
\end{center}
\end{figure}

The Feynman diagrams for the exclusive process
$\chi_{bJ}(Q,q,\lambda)\to
J/\psi(P_1,q_1,\lambda_1)+J/\psi(P_2,q_2,\lambda_2)$ are shown in
Fig. 1.

\subsection{order-$v^2$ QCD amplitude}

 The $Q\bar Q$ analog $\mathcal
{M}_Q(^{2s+1}L_J)$ of the hadronic can be obtained by restricting
the $Q\bar Q$ to have an appropriate spectroscopic state. A given
spin state of the color-singlet $Q\bar Q$ pair can be projected out
by replacing $u(p)\bar{v}(\bar{p})$ or $v(\bar{p})\bar{u}(p)$ with a
projection matrix that can project to a particular spin and color
channel. In our case, the spins of $J/\psi$ and $\chi_{bJ}$ are all
equal to $1$, so the projection matrix can be expressed as
\cite{Braaten:2002fi,Bodwin:2007ga}
%
\begin{subequations}
\begin{eqnarray}
\Pi_3^\mu\epsilon_\mu&=&-\frac{(\not\!p+m_Q)(\not\!P+2E_q)\gamma^\mu(\not\!\bar{p}-m_Q)}
{4\sqrt{2}E_q(E_q+m_Q)}\epsilon_\mu\otimes\frac{\mathbf{1}}{\sqrt{N_c}}\
\ \\
\gamma^0(\Pi_3^\mu\epsilon_\mu)^\dag\gamma^0&=&-\frac{(\not\!\bar{p}-m_Q)\gamma^\mu(\not\!P+2E_q)(\not\!{p}+m_Q)}
{4\sqrt{2}E_q(E_q+m_Q)}\epsilon^*_\mu\otimes\frac{\mathbf{1}}{\sqrt{N_c}},
\label{sprojector}
\end{eqnarray}
\end{subequations}
%
where $E^2_q=P^2/4=m_Q^2-q^2$, $N_c=3$, and \textbf{1} is the unit
color matrix. $\epsilon$ is a spin polarization vector satisfying
$P\cdot\epsilon=0$ and $\epsilon\cdot\epsilon^*=-1$.

After projecting out the $S$-wave color-singlet spin-triplet of $c\bar c$ and the $P$-wave color-singlet spin-triplet of $b\bar b$ , we can expand $\mathcal {M}[b\bar{b}({}^3P_J)\to c\bar{c}({}^3S_1,P_1,\lambda_1)+c\bar{c}({}^3S_1,P_2,\lambda_2)]$ to the order of $\textbf{v}_1^2$ and $\textbf{v}_2^2$, where $\textbf{v}_i^2=\textbf{q}_i^2/m_c^2$. Then we can project out the diagonal, antisymmetric and symmetric traceless components of $b\bar b(^3P_J)$ for $J=0,1,2$ as~\cite{Braaten:2002fi}. It is straightforward to obtain the $Q\bar Q$ analog $\mathcal{M}_{^3P_J}$ as follow:
%
\bqa\label{mquark}
\mathcal {M}_{^3P_0}&=&\frac{i g_s^4 2^6 (N_c^2-1)
M_{J/\psi}^2}{3\sqrt{6} N_c^{3/2} M_{\chi_{b0}}^7}
\left[6M_{\chi_{b0}}^2\epsilon_1^*\cdot\epsilon_2^*-
(\textbf{v}_1^2+\textbf{v}_2^2)
\left[(M_{\chi_{b0}}^2-2M_{J/\psi}^2)\epsilon_1^*\cdot\epsilon_2^*-5P_1\cdot\epsilon^*_2P_2\cdot\epsilon^*_1\right]\right],\nn\\
\mathcal {M}_{^3P_1}&=&0\,,\\
\mathcal {M}_{^3P_2}&=&\frac{i g_s^4 2^{11/2}
(1-N_c^2)\epsilon_{\rho\sigma}}{3 N_c^{3/2}
M_{\chi_{b2}}^7}\left[3M_{\chi_{b2}}^2(
2P_1\cdot\epsilon_2^*P_2^\rho\epsilon_1^{*\sigma} +
2P_2\cdot\epsilon_1^*P_1^\rho\epsilon_2^{*\sigma} -2P_1^\rho
P_2^\sigma\epsilon_1^*\cdot\epsilon_2^*- M_{\chi_{b2}}^2
\epsilon_1^{*\rho}\epsilon_2^{*\sigma})\right.\nn\\
&&\left.+(\textbf{v}_1^2+\textbf{v}_2^2)2M_{J/\psi}^2\left(2P_1^{\rho
} P_2^{\sigma }\epsilon_1^*\cdot \epsilon_2^*+2M_{\chi_{b2}}^2
\epsilon_1^{*\rho }\epsilon_2^{*\sigma }-3\epsilon_2^{*\sigma
}P_1^{\rho }P_2\cdot \epsilon_1^*-3P_1\cdot \epsilon_2^*
\epsilon_1^{*\rho}P_2^\sigma\right)\right]\,,\nn
 \eqa

In Sec.~\ref{factorization}, we give the amplitude in hadronic level $\mathcal{M}[\chi_{bJ}\to J/\psi+J/\psi]$ and the $Q\bar Q$ analog
amplitude $\mathcal {M}[b\bar{b}({}^3P_J)\to c\bar{c}({}^3S_1,P_1,\lambda_1)+c\bar{c}({}^3S_1,P_2,\lambda_2)]$, respectively, in Eqs.~(\ref{nrqcd-exp:CS:ampl}) and (\ref{ampl:ccbar}). Then, we can derive the following relationship from the these two equations, taking $\chi_{b0}$, for example:
%
\bqa
\mathcal
{M}_{\chi_{b0}} =\sqrt{\frac{2M_{J/\psi}\langle\mathcal
{O}_1\rangle_{J/\psi_1}}{2N_c(2E(q_1))^2}}\sqrt{\frac{2M_{J/\psi}\langle\mathcal
{O}_1\rangle_{J/\psi_2}}{2N_c(2E(q_2))^2}}\sqrt{\frac{2M_{\chi_{b0}}\langle\mathcal
{O}_1\rangle_{\chi_{b0}}}{2N_c(2E(Q))^2}}\mathcal {M}_{^3P_0}
\label{relationship} \eqa
\begin{eqnarray}
\langle\mathcal
{O}_1\rangle_{J/\psi}&=&\vert\langle J/\psi|\psi^+\bm{\sigma}\chi|0\rangle\vert^2=\frac{N_c}{2\pi}R^2_{J/\psi}(0)\\
\langle\mathcal{O}_1}\rangle_{\chi_{b0} &=&\frac{1}{3}\vert\langle 0|\chi^+(-\frac{i}{2}\tensor{\bm{D}}\cdot \bm{\sigma})|\chi_{b0}\rangle\vert^2=\frac{3N_c}{2\pi}R^{'2}_{\chi_{b0}}(0)
\end{eqnarray}
As we have discussed, for the precision of our work, we have fixed $2E(q_1)=M_{J/\psi_1}$,
$2E(q_2)=M_{J/\psi_2}$ and $2E(Q)=M_{\chi_{bJ}}$. Again the mass of
$J/\psi$ is identic, so we have
$M_{J/\psi_1}=M_{J/\psi_2}=M_{J/\psi}$.

Substituting Eq.(\ref{mquark}) into Eq.(\ref{relationship}), we reach the order $v^2$ QCD amplitude $\mathcal {M}_{\chi_{bJ}}$:
%
\bqa
\mathcal
{M}^{s}_{\chi_{b0}}&=&\frac{iA_0}{3\sqrt{6}}M_{\chi_{b0}}M_{J/\psi}^2\left\{-6M_{\chi_{b0}}^2\epsilon_1^*\cdot\epsilon_2^*+
(\textbf{v}_1^2+\textbf{v}_2^2)
\left[(M_{\chi_{b0}}^2-2M_{J/\psi}^2)\epsilon_1^*\cdot\epsilon_2^*-5P_1\cdot\epsilon^*_2P_2\cdot\epsilon^*_1\right]\right\}\,,\nn\\
\mathcal {M}^{s}_{\chi_{b1}}&=&0\,,\\
\mathcal
{M}^{s}_{\chi_{b2}}&=&\frac{iA_2\sqrt{2}}{6}M_{\chi_{b2}}\epsilon_{\rho\sigma}\left\{3M_{\chi_{b2}}^2(
2P_1\cdot\epsilon_2^*P_2^\rho\epsilon_1^{*\sigma} +
2P_2\cdot\epsilon_1^*P_1^\rho\epsilon_2^{*\sigma} -2P_1^\rho
P_2^\sigma\epsilon_1^*\cdot\epsilon_2^*- M_{\chi_{b2}}^2
\epsilon_1^{*\rho}\epsilon_2^{*\sigma})\right.\nn\\
&&\left.+(\textbf{v}_1^2+\textbf{v}_2^2)2M_{J/\psi}^2\left(2P_1^{\rho
} P_2^{\sigma }\epsilon_1^*\cdot \epsilon_2^*+2M_{\chi_{b2}}^2
\epsilon_1^{*\rho }\epsilon_2^{*\sigma }-3\epsilon_2^{*\sigma
}P_1^{\rho }P_2\cdot \epsilon_1^*-3P_1\cdot \epsilon_2^*
\epsilon_1^{*\rho}P_2^\sigma\right)\right\}\,,\nn
\label{strongamplitude} \eqa
 where
 \bqa
A_0&=&\frac{-g^4_s2^6(N_c^2-1)}{N_c^{3}M_{J/\psi}M_{\chi_{b0}}^{17/2}}\langle
J/\psi_1|\psi^\dag
\bm{\sigma}\cdot\bm{\epsilon}(\lambda_1)\chi|0\rangle\langle
J/\psi_2|\psi^\dag
\bm{\sigma}\cdot\bm{\epsilon}(\lambda_2)\chi|0\rangle
\frac{1}{\sqrt{3}}\langle0|\chi^\dag (-\tfrac{i}{2}
\tensor{\bm{D}}\cdot \bm{\sigma}\ )\psi|\chi_{b0}\rangle\\
A_2&=&\frac{-g^4_s2^6(N_c^2-1)}{N_c^{3}M_{J/\psi}M_{\chi_{b2}}^{17/2}}\langle
J/\psi_1|\psi^\dag
\bm{\sigma}\cdot\bm{\epsilon}(\lambda_1)\chi|0\rangle\langle
J/\psi_2|\psi^\dag
\bm{\sigma}\cdot\bm{\epsilon}(\lambda_2)\chi|0\rangle
\sum_{ij}\langle0|\chi^\dag (-\tfrac{i}{2}
\tensor{{D}}^{(i}{\sigma}^{j)} \epsilon^{ij}(\lambda)\nn
)\psi|\chi_{b2}\rangle
\eqa
%

\subsection{Helicity Amplitude}
The polarized decay width and branching ratios can offer more useful information for both experimentation and theory, which are lost in the unpolarized ones. As we have gotten the amplitude of $\mathcal {M}[\chi_{bJ}\to J/\psi J/\psi]$, for $J=0,1,2$, we can easily know the helicity amplitude by helicity amplitude formalism~\cite{Jacob:1959at}. According to Ref.~\cite{Haber:1994pe}, we can obtain the corresponding helicity
amplitude:
%
\bqa
\mathcal {M}^{(J)}_{\lambda_1\,\lambda_2;\,\mu}=\tilde{\mathcal
{M}}_{\lambda_1\lambda_2}e^{i\mu\varphi}d^{(J)}_{m,\,\lambda_1-\lambda_2}(\theta)\,,
\eqa
%
where $\lambda_1$, $\lambda_2$ is the helicity of $J/\psi$, $\tilde{\mathcal
{M}}_{\lambda_1\lambda_2}$ is the reduced helicity amplitude which is a function of J, $\lambda_1$, $\lambda_2$ and particle masses but is independent of $\theta$ and $\varphi$, $\mu$ is
the $\chi_{bJ}$ spin projection on fixed axe, and $\theta$ and $\varphi$
are polar and azimuthal angles of one of the final $J/\psi$ in the
$\chi_{bJ}$ rest frame. We can obtained
the reduced helicity amplitudes $\tilde{\mathcal {M}}$ as follows:\\
For$\chi_{b0}$,
\bqa
\tilde{\mathcal {M}}_{0,\,0}^s(\chi_{b0})&=&
\frac{iA_0}{12\sqrt{6}}M_{\chi_{b0}}\left[12M_{\chi_{b0}}^2(2M_{J/\psi}^2-M_{\chi_{b0}}^2)+(\textbf{v}_1^2+\textbf{v}_2^2)
(8M_{J/\psi}^4-3M_{\chi_{b0}}^4+12M_{J/\psi}^2M_{\chi_{b0}}^2)\right]\,,\nn\\
\tilde{\mathcal {M}}_{1,1}^s(\chi_{b0})
&=&\frac{iA_0}{3\sqrt{6}}M_{\chi_{b0}}M_{J/\psi}^2\left[6M_{\chi_{b0}}^2+(\textbf{v}_1^2+\textbf{v}_2^2)
(2M_{J/\psi}^2-M_{\chi_{b0}}^2)\right]\,,
\eqa
For $\chi_{b2}$,
\bqa
\tilde{\mathcal {M}}_{0,\,0}^s(\chi_{b2})&=&
\frac{iA_2}{6\sqrt{3}}M_{\chi_{b2}}\left[3M_{\chi_{b2}}^2(4M_{J/\psi}^2+M_{\chi_{b2}}^2)-({\textbf{v}_1^2}
+{\textbf{v}_2^2})2M_{J/\psi}^2(3M_{\chi_{b2}}^2+4M_{J/\psi}^2)\right].\nn\\
 \tilde{\mathcal {M}}_{1,1}^s(\chi_{b2})
&=&
\frac{iA_2}{3\sqrt{3}}M_{\chi_{b2}}\left[6M_{\chi_{b2}}^2M_{J/\psi}^2-({\textbf{v}_1^2}
+{\textbf{v}_2^2} )M_{J/\psi}^2(M_{\chi_{b2}}^2+4M_{J/\psi}^2)\right]\,,\nn\\
\tilde{\mathcal {M}}_{1,\,0}^s(\chi_{b2}) &=&
\frac{iA_2}{12}M_{J/\psi}M_{\chi_{b2}}^2\left[12M_{\chi_{b2}}^2-({\textbf{v}_1^2}
+{\textbf{v}_2^2}
)(M_{\chi_{b2}}^2+12M_{J/\psi}^2)\right]\,,\nn\\
\tilde{\mathcal {M}}_{1,-1}^s(\chi_{b2}) &=&
\frac{iA_2\sqrt{2}}{6}M_{\chi_{b2}}^3\left[3M_{\chi_{b2}}^2-4({\textbf{v}_1^2}
+{\textbf{v}_2^2})M_{J/\psi}^2 \right]\,,
\eqa

According to parity invariance, there are only two independent helicity amplitudes for $\chi_{b0}$ and four independent helicity amplitudes for $\chi_{b2}$. The unpolarized amplitude squared can be obtained by integrating the polar angle and summing all the helicity states. The relationship between the unpolarized amplitude squared and the
helicity amplitude squared is:
\begin{subequations}
 \bqa |\mathcal
{M}^s_{\chi_{b0}}|^2&=&|\tilde{\mathcal
{M}}_{0,\,0}^s(\chi_{b0})|^2+2|\tilde{\mathcal
{M}}_{1,\,1}^s(\chi_{b0})|^2\\
|\mathcal {M}^s_{\chi_{b2}}|^2&=&|\tilde{\mathcal
{M}}_{0,\,0}^s(\chi_{b2})|^2+2|\tilde{\mathcal
{M}}_{1,\,1}^s(\chi_{b2})|^2
 +4|\tilde{\mathcal {M}}_{1,\,0}^s(\chi_{b2})|^2
 +2|\tilde{\mathcal {M}}_{1,\,-1}^s(\chi_{b2})|^2
\eqa
\end{subequations}

Here we give some general prediction of these helicity amplitudes by hadron helicity selection rule. In this exclusive decay process, when we consider the lowest-order strong interaction in the limit $m_b\to \infty$ with $m_c$ fixed, we can obtain the asymptotic behavior for ${\rm Br}[\chi_{bJ} \to J/\psi J/\psi]$:
\bqa
{\rm Br_{\rm \, str}} [\chi_{bJ}\to
J/\psi(\lambda_1)+J/\psi(\lambda_2)] &\sim& \alpha_s^2 \,v_c^6
\left(m_c^2\over m_b^2\right)^{2+|\lambda_1+\lambda_2|}\,,
\label{Brstr:scaling1}
\eqa
where $v_c$ is the relative velocity of $c\bar c$ in $J/\psi$, and the factor $v^6$ comes from the wave function at the origin of $J/\psi$.

From Eq.~(\ref{Brstr:scaling1}), we can see clearly that the helicity configurations of two $J/\psi$ decide the scaling behavior of the branching ratio. We figure out that when the decay configuration is $\lambda_1+\lambda_2=0$, i.e.\ ($\lambda_1$, $\lambda_2$)=(0, 0), and ($\lambda_1$, $\lambda_2$)=(1, -1), the branching ration exhibits the slowest asymptotic decrease ${\rm Br_{str}}\sim 1/m_b^4$. And we can also expect ${\rm Br_{str}}\sim 1/m_b^6$ when ($\lambda_1$, $\lambda_2$)=(0, 1) and the ${\rm Br_{str}}\sim 1/m_b^8$ when ($\lambda_1$, $\lambda_2$)=(1, 1). In fact, due to the nonzero charm mass helicity conservation is violated.

In particular, the helicity state $(0,0)$ of the channel $\chi_{b1} \to J/\psi J/\psi$ is zero because it is strictly forbidden due to the conflict
between parity and angular momentum conservation. We call this kind of process an unnatural decay process ~\cite{Chernyak:1983ej}.

\subsection{Electromagnetic amplitude}

For completeness, we also consider the electromagnetic contributions
to $\chi_{bJ}\to J/\psi J/\psi$. There are four QED diagrams, two of
which have the same topology as Fig.1, but with gluons replaced by
photons, lead to the amplitude that has the same form as
Eq.~(\ref{strongamplitude}) except $\alpha_s^2$ is replaced by
$e_b^2e_c^2\alpha^2$. But their contributions are much more
suppressed than those from the fragmentation diagrams in Fig.2, and we
will not consider them.

We can see that in Fig.2, both $J/\psi$ decay from the corresponding virtual photon, so we can use their decay constant instead of the vacuum matrix element to describe them. With the definition given in Ref.~\cite{Hwang:1997ie},
\begin{eqnarray}
\left\langle 0 \right\vert \bar{c} \gamma^\mu c \left\vert J/\psi (P,\epsilon) \right\rangle = f_{J/\psi}M_{J/\psi} \epsilon^\mu \,
\end{eqnarray}
where $P$ is the momentum of $J/\psi$, $\epsilon$ is its polarization vector, and $f_{J/\psi}$ is the so-called decay constant.
\begin{figure}[h]
\begin{center}

\includegraphics[scale=0.8]{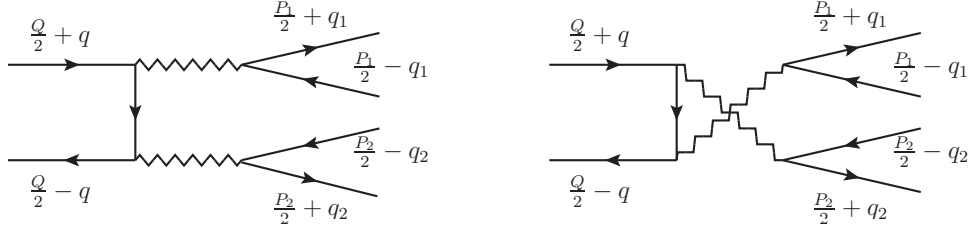}
\caption{ Lowest-order QED diagrams that contributes to $\chi_{bJ}
\to J/\psi+J/\psi$. Only the fragmentation-type diagrams are
retained, whereas the other two, which can be obtained by replacing
the gluons in Fig.1 with photons, have been suppressed.
\label{feynman:diag:gamma}}
\end{center}
\end{figure}

We can infer the QED fragmentation contribution to the amplitude in
Fig.~\ref{feynman:diag:gamma};
in the following results, we have already included all order relativistic corrections by using the decay constant:
%
\begin{subequations}
\bqa
\mathcal {M}^{em}_{\chi_{b0}}&=&
\frac{i 2 B_0 }{\sqrt{6}}\frac{M_{J/\psi}^2}{M_{\chi_{b0}}}\left[(3M_{\chi_{b0}}^2-8M_{J/\psi}^2)\epsilon_1^*\cdot\epsilon_2^*M_{\chi_{b0}}^2+
2(2M_{J/\psi}^2-3M_{\chi_{b0}}^2)P_1\cdot\epsilon^*_2P_2\cdot\epsilon^*_1\right]\,,\\
\mathcal
{M}^{em}_{\chi_{b1}}&=&8B_1\frac{M_{J/\psi}^4}{M_{\chi_{b1}}^2}\epsilon_{\alpha\beta\rho\sigma}Q^\alpha\epsilon^\beta
({P_1}^{\sigma } \epsilon_1^{*\rho } P_1\cdot \epsilon_2^*+P_1^{\rho
} \epsilon_2^{*\sigma
   } P_2\cdot
  \epsilon_1)\,,\\
\mathcal {M}^{em}_{\chi_{b2}}&=&i 2 B_2
\sqrt{2}M_{J/\psi}^2M_{\chi_{b2}}\epsilon_{\rho\sigma} \\&&
[2\epsilon_1^{*\rho } P_2^{\sigma }P_1\cdot
\epsilon_2^*+2\epsilon_2^{*\rho } P_1^{\sigma } P_2\cdot
\epsilon_1^*
+ \left( 2M_{J/\psi}^2-m_b^2\right) \epsilon_2^{*\rho }
   \epsilon_1^{*\sigma }+2P_1^\rho P_1^\sigma
   \epsilon_1^*\cdot \epsilon_2^*]\,,\nn
\eqa
\end{subequations}
%
where
%
\begin{subequations}
\bqa
B_0&=&-e_b^2e_c^2e^4 f^2_{J/\psi}
\frac{4(M_{\chi_{b0}}^2-2M_{J/\psi}^2)^{-2}}{\sqrt{M_{\chi_{bJ}}}\,M_{J/\psi}^4}
\frac{1}{\sqrt{3}}\langle0|\chi^\dag (-\tfrac{i}{2}
\tensor{\bm{D}}\cdot \bm{\sigma}\ )\psi|\chi_{b0}\rangle\,,\\
B_1&=&-e_b^2e_c^2e^4 f^2_{J/\psi}
\frac{4(M_{\chi_{b1}}^2-2M_{J/\psi}^2)^{-2}}{\sqrt{M_{\chi_{bJ}}}\,M_{J/\psi}^4}
\frac{1}{\sqrt{2}}\langle0|\chi^\dag (-\tfrac{i}{2}
\tensor{\bm{D}}\times \bm{\sigma}\cdot\bm{\epsilon}(\lambda)
)\psi|\chi_{b1}\rangle\,,\\
B_2&=&-e_b^2e_c^2e^4 f^2_{J/\psi}
\frac{4(M_{\chi_{b2}}^2-2M_{J/\psi}^2)^{-2}}{\sqrt{M_{\chi_{bJ}}}\,M_{J/\psi}^4}
\sum_{ij}\langle0|\chi^\dag (-\tfrac{i}{2}
\tensor{{D}}^{(i}{\sigma}^{j)} \epsilon^{ij}(\lambda)
)\psi|\chi_{b2}\rangle\,.
\eqa \end{subequations}
%
Following are the helicity amplitudes:
\begin{subequations}
\bqa
\tilde{\mathcal {M}}_{0,\,0}^{em}(\chi_{b0})&=&iB_0\ \
\frac{16}{\sqrt{6}}M_{\chi_{b0}}M_{J/\psi}^4\,,
\tilde{\mathcal {M}}_{1,1}^{em}(\chi_{b0})=iB_0
\frac{4}{\sqrt{6}}M_{\chi_{b0}}M_{J/\psi}^2(8M_{J/\psi}^2-3M_{\chi_{b0}}^2)\,,\\
\tilde{\mathcal {M}}_{1,\,0}^{em}(\chi_{b1})&=&iB_1
4M_{J/\psi}^3(4M_{J/\psi}^2-M_{\chi_{b1}}^2)\,,\\
\tilde{\mathcal {M}}_{0,\,0}^{em}(\chi_{b2})&=&iB_2\ \ \
\frac{16}{\sqrt{3}}M_{J/\psi}^4M_{\chi_{b2}}\,,
\tilde{\mathcal {M}}_{1,1}^{em}(\chi_{b2}) =iB_2
\frac{8}{\sqrt{3}}M_{J/\psi}^4M_{\chi_{b2}}\,,\\
\tilde{\mathcal {M}}_{1,\,0}^{em}(\chi_{b2})&=& iB_2 4 M_{\chi_{b2}}^2M_{J/\psi}^3\,,\ \ \ \
\tilde{\mathcal {M}}_{1,-1}^{em}(\chi_{b2})=iB_2
4\sqrt{2}M_{J/\psi}^2M_{\chi_{b2}}\left(M_{\chi_{b2}}^2-2
M_{J/\psi}^2\right)\,,\nonumber \eqa
\end{subequations}

The relationship between the unpolarized amplitude squared and the
helicity amplitude squared is similar to the relation in the QCD section except a new equation for $\chi_{b1}$:
\bqa
|\mathcal {M}^{em}_{\chi_{b1}}|^2&=&4|\tilde{\mathcal
{M}}_{1,\,0}^{em}(\chi_{b1})|^2
 \eqa

\subsection{Decay width}
As we have known the amplitude squared, we can easily derive the
decay width of $\chi_{bJ}\to J/\psi J/\psi$. The polarized decay
width can be expressed as
%
\bqa
\Gamma_{\lambda_1,\,\lambda_2}=\frac{1}{2M_{\chi_{bJ}}}\times\frac{1}{2}\times\Phi_2\int^1_{-1}\frac{d\cos\theta}{2}|\mathcal
{M}^{(J)}_{\lambda_1\,\lambda_2;\,\mu}|^2
\label{decay1} \eqa
%
The unpolarized decay width can be expressed as
%
\bqa
\Gamma_{unp}=\frac{1}{2M_{\chi_{bJ}}}\times\frac{1}{2}\times\Phi_2\int^1_{-1}\frac{d\cos\theta}{2}\sum_{\lambda_1\,\lambda_2}|\mathcal
{M}^{(J)}_{\lambda_1\,\lambda_2;\,\mu}|^2
\label{decay2} \eqa
%
where $|\mathcal {M}^{(J)}_{\lambda_1\,\lambda_2;\,\mu}|^2=|\mathcal
{M}^{(J)}_{\lambda_1\,\lambda_2;\,\mu\,s}-\mathcal
{M}^{(J)}_{\lambda_1\,\lambda_2;\,\mu\,em}|^2$, $\Phi_2=\frac{1}{8\pi}
\sqrt{1-\frac{4 M_{J/\psi}^2}{M_{\chi_{bJ}}^2}}$.
\\

\section{Numerical analysis} \label{numericalresult}
\subsection{Input parameters}
In this section, we will apply the results obtained in Sec.~\ref{fullqcd} to give some phenomenological predictions. Before we carry out the numerical calculations, we need to fix several input parameters, such as $M_{\chi_{bJ}}$, $M_{J/\psi}$, $\alpha$, $\alpha_s$, $\langle \mathcal{O}_1\rangle_{J/\psi}$, $\langle \mathcal{O}_1\rangle_{\chi_{bJ}}$, $f_{J/\psi}$ and $\langle v^2\rangle_{J/\psi}$.

First, we need to fix the values of coupling constants; in our work, we set the running QCD strong coupling constant $\alpha_s(M_{\chi_{bJ}}/2)=0.22$ by using the two-loop formula with $\Lambda_{\overline{\rm MS}}=0.338$ GeV \cite{Zhang:2005cha, Gong:2007db} and the electromagnetic fine structure constant $\alpha=1/137$.

Next,The NRQCD matrix element $\langle\mathcal {O}_1\rangle_{\chi_{bJ}}$ can be obtained from the derivative of radial wave functions near the origin in the potential model, for P-wave is $\langle\mathcal
{O}_1\rangle_{\chi_{bJ}}\approx\frac{3N_c}{2\pi}|R^\prime_{1P}(0)|^2$ as in Ref.\cite{Braaten:2002fi,Bodwin:2007zf}. In Ref.\cite{Eichten:1995ch} the values of $|R^\prime_{1P}(0)|^2$ for four potentials have been listed; we use the value of the Cornell potential, and then obtain $\langle\mathcal {O}_1\rangle_{\chi_{bJ}}=2.96\ \mathrm{GeV}^5$. The color-singlet NRQCD matrix elements of $J/\psi$, i.e.$\langle\mathcal {O}_1\rangle_{J/\psi}$, can be obtained from the electromagnetic decay rate of the $J/\psi$ in which the $\alpha_s$ leading order is considered. Using the measured dielectric width $5.5$ KeV, we obtained $\langle\mathcal {O}_1\rangle_{J/\psi}=0.268\ \mathrm{GeV}^3$.

The values for the physical masses of the involving hadrons are taken from Ref.\cite{pdg} as $M_{\chi_{b0}}=9.859\ \mathrm{GeV}$, $M_{\chi_{b1}}=9.893\ \mathrm{GeV}$, $M_{\chi_{b2}}=9.912\ \mathrm{GeV}$, $M_{J/\psi}=3.097\ \mathrm{GeV}$. The decay constant for $J/\psi$ is taken as $f_{J/\psi}$ = $0.406$ GeV as in Ref.~\cite{Hwang:1997ie}.

Considering the order $v^2$ relativistic corrections are calculated, we also need to know the value of $\langle v^2\rangle_{J/\psi}$. Here we adopt the value $\langle v^2\rangle_{J/\psi_1}=\langle v^2\rangle_{J/\psi_2}=0.225$ extracted from the recent Cornell-potential-model-based analysis in Ref.\cite{:2008vj}.

\subsection{The decay width of $\chi_{bJ}\to J/\psi J/\psi$}
With the input parameters fixed, we present the numerical results of the polarized and unpolarized decay widths for $\chi_{bJ}\to J/\psi J/\psi$ in Table~\ref{table1}.

\begin{table}[h]
\caption{\label{table1}%
The polarized and unpolarized decay widths for $\chi_{bJ}\to J/\psi J/\psi$, where we list the leading-order QCD contributions of $\Gamma(\chi_{bJ}\to J/\psi J/\psi)$, QED corrections, $v^2$-order QCD corrections (including LO contributions) and both the QED and QCD corrections. The values are all in units of eV.}
\begin{ruledtabular}
\begin{tabular}{ccccccc}
  &  &$\Gamma_{0,\ 0}$&$\Gamma_{1,1}$   & $\Gamma_{1,\ 0}$   &  $\Gamma_{1,-1}$  &$\Gamma_{unp}$ \,\\
\hline
\multirow{3}{*}{LO}&
$\chi_{b0}$     &    4.334& 0.262 & ---&--- &4.859\\
&$\chi_{b1}$     &  ---   & ---  &--- & ---& ---\\
&$\chi_{b2}$     &    1.251&0.099&0.758&3.881&12.240\\
\hline
\multirow{3}{*}{QED}&
$\chi_{b0}$  &   4.338& 0.266&---                 & ---& 4.870\\
&$\chi_{b1}$  &   ---  & ---  &$2.26\times10^{-7}$&--- & $9.04\times10^{-7}$\\
&$\chi_{b2}$  &   1.250& 0.098&0.757               &3.872&12.216\\
\hline
\multirow{3}{*}{$v^2$-QCD}&
$\chi_{b0}$    &   5.067& 0.231&--- &--- & 5.530\\
&$\chi_{b1}$    &  ---   & ---  &---& ---&--- \\
&$\chi_{b2}$    &    1.078& 0.079&0.639&3.439&10.673\\
\hline
\multirow{3}{*}{QED$\&$QCD}&
$\chi_{b0}$  &    5.070& 0.235&--- & ---& 5.541\\
&$\chi_{b1}$  &  ---   & ---  &$2.26\times10^{-7}$&--- & $9.04\times10^{-7}$\\
&$\chi_{b2}$  &    1.077& 0.079&0.638&3.431&10.651\\
\end{tabular}
\end{ruledtabular}

\end{table}

 We can see that the relative order-$v^2$ contributions are very prominent. It is found to increase the lower-order prediction for the decay width by about $13.8\%$ for $\chi_{b0}\to J/\psi J/\psi$, and about $12.8\%$ for $\chi_{b2}\to J/\psi J/\psi$. The QED contributions are very small, only about $0.2\%$ for $\chi_{b0}\to J/\psi J/\psi$ and $\chi_{b2}\to J/\psi J/\psi$.

 We can also see in Table~\ref{table1} that the polarized decay width is consistent with the helicity selection rule in general. As we analyzed in Sec.~\ref{fullqcd}, the configurations of two $J/\psi$ with ($\lambda_1$, $\lambda_2$)=(0, 0) and ($\lambda_1$, $\lambda_2$)=(1, -1) bear the smallest suppression and contribute the largest component to total decay width. The ($\lambda_1$, $\lambda_2$)=(0, 1) channel gives the second largest contribution and the contribution from the ($\lambda_1$, $\lambda_2$)=(1, 1) channel is the smallest one. So we speculate the most observable channels are ($\lambda_1$, $\lambda_2$)=(1, -1) for $\chi_{b2}$ and ($\lambda_1$, $\lambda_2$)=(0, 0) for $\chi_{b0}$ respectively.

\subsection{The decay branching ratio of $\chi_{bJ}\to J/\psi J/\psi$}
Nevertheless, what we are more interested in is the branching ratio, so we need to know the total decay width. According to Ref.~\cite{Bodwin:1994jh}, the decay rate of $\chi_{bJ}$ into light hadrons (LH) is
%
\bqa
\Gamma(\chi_{bJ}\to
\mathrm{LH})=\frac{2\mathrm{Im}\,f_1({}^3P_J)}{m_b^4}\langle\mathcal
{O}_1({}^3P_J)\rangle_{\chi_{bJ}}+\frac{2\mathrm{Im}\,f_8({}^3S_1)}{m_b^2}\langle\mathcal
{O}_8({}^1S_0)\rangle_{\chi_{bJ}}\,.
\eqa
%

The imaginary parts of $f_1({}^3P_J)$ and $f_8({}^1S_0)$ have been
calculated up to order $\alpha_s^3$ in Ref.~\cite{Petrelli:1997ge},
but for $f_1({}^3P_0)$ and $f_1({}^3P_2)$ different results were given
in Ref.\cite{Barbieri:1980yp}, so we take the leading-order results
as
%
\begin{subequations}
\bqa
\Gamma_{tot}(\chi_{b0})&=&\frac{64\pi\alpha_s^2\langle\mathcal
{O}_1({}^3P_J)\rangle_{\chi_{bJ}}}{3M_{\chi_{b0}}^4}\left(1+\frac{n_f\,M_{\chi_{b0}}^2\langle\mathcal
{O}_8({}^1S_0)\rangle_{\chi_{bJ}}}{16\langle\mathcal
{O}_1({}^3P_J)\rangle_{\chi_{bJ}}}\right),\\
\Gamma_{tot}(\chi_{b1})&=&\frac{64\pi\alpha_s^2\langle\mathcal
{O}_1({}^3P_J)\rangle_{\chi_{bJ}}}{3M_{\chi_{b1}}^4}\left(0+\frac{n_f\,M_{\chi_{b1}}^2\langle\mathcal
{O}_8({}^1S_0)\rangle_{\chi_{bJ}}}{16\langle\mathcal
{O}_1({}^3P_J)\rangle_{\chi_{bJ}}}\right),\\
\Gamma_{tot}(\chi_{b2})&=&\frac{64\pi\alpha_s^2\langle\mathcal
{O}_1({}^3P_J)\rangle_{\chi_{bJ}}}{3M_{\chi_{b2}}^4}\left(\frac{4}{15}+\frac{n_f\,M_{\chi_{b2}}^2\langle\mathcal
{O}_8({}^1S_0)\rangle_{\chi_{bJ}}}{16\langle\mathcal
{O}_1({}^3P_J)\rangle_{\chi_{bJ}}}\right),
\eqa
\end{subequations}
%
where $n_f=4$. According to Ref.~\cite{Bodwin:2007zf}, we take
$\frac{\langle\mathcal
{O}_8({}^1S_0)\rangle_{\chi_{bJ}}}{\langle\mathcal
{O}_1({}^3P_J)\rangle_{\chi_{bJ}}}=0.0021\,\mathrm{GeV}^{-2}$, then
we can obtain the value of total decay.

Another method is to calculate the decay width of E1 transitions of $\chi_{bJ}\to
\gamma\Upsilon$, according to the branching ratio of $\chi_{bJ}\to
\gamma\Upsilon$ from Ref.\cite{Amsler:2008zzb}, we can obtain
the total decay width of $\chi_{bJ}$. The E1 transitions width is
defined as \cite{Brambilla:2004wf}
%
\bqa
\Gamma(1{}^3P_J\xrightarrow[{}]{E1} 1{}^3S_1+\gamma)=\frac{4\alpha
e_Q^2}{3}\frac{1}{3}\,k^3\left|\frac{3}{k}\int^\infty_0r^2drR_{11}(r)R_{10}(r)
\left[\frac{kr}{2}j_0(\frac{kr}{2})-j_1(\frac{kr}{2})\right]\right|^2
\eqa
%
where $k$ is the photon momentum, and $R_{nl}(r)$ is radial wave-function.

The total decay widths of $\chi_{bJ}$ obtained from the two methods are listed in Table~\ref{table5}.

\begin{table}[h]
\caption{\label{table5}%
Total decay width of $\chi_{bJ}$. One is obtained by E1 transitions, which is from Ref.~\cite{Brambilla:2004wf} and PDG data. Another is from the $\chi_{bJ}$ decay into LH in leading order. The values are all in units of KeV.}
\begin{ruledtabular}
\begin{tabular}{lcc}

  & ${\rm E1}$ &${\rm LH}$\\
\hline
$\Gamma(\chi_{b0})$        &    $>319$& 1068 \\
$\Gamma(\chi_{b1})$        &  69   & 52  \\
$\Gamma(\chi_{b2})$        &    124& 317
\end{tabular}
\end{ruledtabular}
\end{table}

\begin{table}[h]
\caption{\label{table6}%
  Total decay width and branching ratio of $\chi_{bJ}\to J/\psi J/\psi$. Here we give the total decay width including QED and order-$v^2$ relativistic corrections. For comparison, we list the results obtained in~\cite{Braguta:2010zz} within NRQCD. We also show the branching ratios obtained by $\chi_{bJ}$ decay into light hadrons and E1 transition.
}
\begin{ruledtabular}
\begin{tabular}{lcccc}[h]
& $\Gamma$(eV) & $\Gamma$\cite{Braguta:2010zz}(eV)&$\mbox{Br}_{\rm E1}(10^{-5})$ &$\mbox{Br}_{\rm LH}(10^{-5})$      \,\\
\hline
$\chi_{b0}\to J/\psi J/\psi$     &  $ 5.54   $             &$27^{+5}_{-2.5}\pm19\pm13$   &$<1.7$                & $0.5$ \\
$\chi_{b1}\to J/\psi J/\psi$     &  $ 9.04\times 10^{-7}$  &--                           &$1.4\times10^{-6}$    & $1.9\times10^{-6}$  \\
$\chi_{b2}\to J/\psi J/\psi$     &  $ 10.6   $             &$65^{+14}_{-12}\pm46\pm32$     &$8.6$                 & $3.4$\\
\end{tabular}
\end{ruledtabular}
\end{table}


From all these values we can obtain the branching ratio in Table~\ref{table6}. For comparison, we juxtapose the decay width of $\chi_{bJ}\to J/\psi J/\psi$ from our calculation and the results obtained in~\cite{Braguta:2010zz} within NRQCD, where the first errors come from the uncertainties in the distribution for final-state mesons, the second in $\alpha_s$, and the last errors are associated with power-law corrections.

We can see that the results obtained in~\cite{Braguta:2010zz} are in general larger than ours.
We have compared our analytical expressions with \cite{Braguta:2010zz} and found a difference by a factor of 2 in the formula of decay rate which will reduce the result of \cite{Braguta:2010zz} half. In addition, the various input parameters such as NRQCD matrix elements, $\alpha_s$, and particle masses can bring the large uncertainties. Finally, because we expand the pole mass $m_c$ around the physical mass $M_{J/\psi}$ instead of the traditional matching method, this will decrease our predictions comparing to the orthodox method. Although the results seem to have a sensible difference, we believe the order of magnitude is correct.

\subsection{Some predictions of observation potential for $\chi_{bJ}\to J/\psi J/\psi$}
We use the branching ratio given in the Table~\ref{table6} to explore the possibility for these processes to be observed in experiments.
Considering the branching ratio for each of the decays $J/\psi\to \mu^+
\mu^-$ is about 6\%, we find the branching ratio of $\chi_{bJ}$ decay to muon pair is $ {\rm Br}[\chi_{b0}\to J/\psi J/\psi \to 4 \mu]
\approx 0.6\times 10^{-7}$ and $ {\rm Br}[\chi_{b2}\to J/\psi J/\psi
\to 4 \mu] \approx 3.1\times 10^{-7}$. The total cross section for $\chi_{b0}$ and $\chi_{b2}$ production
at Tevatron energy according to \cite{Braguta:2005gw} is
 \bqa
 \sigma(p\bar{p}\to \chi_{b0}+X )&=&250\, \mathrm{n b},\nn\\
 \sigma(p\bar{p}\to \chi_{b2}+X )&=&320\, \mathrm{n b},
\eqa
We estimate there are about $50\sim150$ for $\chi_{b0}$ and $400\sim1000$ for
$\chi_{b2}$ produced events in Tevatron Run 2 that achieve an integrated luminosity of about 10 ${\rm
fb}^{-1}$ by April 2011.

Similarly, we can combine the cross sections at LHC, which are about 6 times larger than the corresponding cross sections
at Tevatron~\cite{Braguta:2005gw}, to predict the number of produced
events that may reach between 1500 and 4500 for $\chi_{b0}$ and between
12000 and 30000 for $\chi_{b2}$ with the accumulated luminosity 50 fb$^{-1}$ of LHC in 2010, and with the acceptance and efficiency of detector considered, we expect $15\sim 45$ and $120\sim 300$ observed events per year for $\chi_{b0}$ and $\chi_{b2}$ respectively .

\subsection{The decay width and branching ratio of $\chi_{cJ}\to VV(V\to\omega,\phi)$}
The constituent quark model, which treats the light mesons as
non-relativistic bound states, is also frequently invoked as an
alternative method for a quick order-of-magnitude estimate.  In this
sense, the preceding formulas derived for $\chi_{bJ}\to J/\psi
J/\psi$ can be applied to describe the decay processes $\chi_{cJ}
\to VV$, once we understand that we are working with the constituent
quark model.

\begin{table}[h]
\caption{\label{table7}%
  Parameters used for numerical calculation in $\chi_{cJ}\to \phi \phi$ and $\chi_{cJ}\to \omega \omega$
}
\begin{ruledtabular}
\begin{tabular}{lccccccc}
 &$M_V$(GeV)   & $\alpha_s$   &  $|R_{sV}(0)|^2$(GeV${}^3$)  &$|R^\prime_{\chi_c} |^2$(GeV${}^5$) &$M_{\chi_{c0}}$(
 GeV)&$M_{\chi_{c1}}$(GeV)&$M_{\chi_{c2}}$(GeV)\,\\
\hline
$\chi_{cJ}\to \omega \omega$       & 0.78265  &0.3&0.11 & 0.075&$3.41475$ &$3.51066$ &$3.55620$\,\\
$\chi_{cJ}\to \phi \phi$           &
1.019455&0.3&0.19&0.075&$3.41475$ & $3.51066$ &$3.55620$
\end{tabular}
\end{ruledtabular}
\end{table}

\begin{table}[h]
\caption{\label{table8}%
  Decay widths and Branching ratios of $\chi_{cJ}\to \phi \phi$ and $\chi_{cJ}\to \omega \omega$. The second and third rows are the results of our calculation. The next two rows are the results obtained in~\cite{Braguta:2005gw} within ``$\phi_3$'' model. The next row is the branching fractions obtained by BESIII~\cite{Ablikim:2011}, where we take the results of combined final state. The last row is the PDG~\cite{pdg} results. In $\chi_{cJ}\to \omega \omega$, there are only our results, BESIII results and PDG results.}
\begin{ruledtabular}
\begin{tabular}{lccc
}
& $\chi_{c0}\to \phi \phi$  &$\chi_{c1}\to \phi \phi$ &$\chi_{c2}\to \phi \phi$ \,\\
\hline
$\Gamma$(keV)                                 & 3.3 & $1.9\times10^{-7}$ & 5.9\\
Br($10^{-4}$)                              & 3.2 & $2.2\times10^{-6}$ & 30\\
$\Gamma$(keV)~\cite{Braguta:2005gw}          & 2.10 & --                 & 3.38\\

Br($10^{-4}$)~\cite{Braguta:2005gw}        & 3.01 & --                & 21.3\\
Br($10^{-4}$)~\cite{Ablikim:2011}(BESIII)     & $8.0\pm 0.3\pm 0.8$ & $4.4\pm 0.2\pm 0.5$ & $10.7\pm 0.3\pm 1.2$\\
Br($10^{-4}$)~\cite{pdg}(PDG)      & $9.2\pm 1.9$ & -- & $ 14.8\pm 2.8$\\
\hline
& $\chi_{c0}\to \omega \omega$  & $\chi_{c1}\to \omega \omega$  &$\chi_{c2}\to \omega \omega$ \,\\
\hline
$\Gamma$(keV)          & 2.3  &$2.2\times10^{-7}$ & 3.2\\
Br($10^{-4}$)       & 2.2  &$2.6\times10^{-6}$ & 16\\
Br($10^{-4}$)~\cite{Ablikim:2011}(BESIII)      & $9.5\pm 0.3\pm 1.1$ & $6.0\pm 0.2\pm 0.7$ & $8.9\pm 0.3\pm 1.1$\\
Br($10^{-4}$)~\cite{pdg}(PDG)       & $22\pm 7$ & -- & $19.0\pm 6.0$
\end{tabular}
\end{ruledtabular}
\end{table}

We have listed the numerical values for parameters in Table.
\ref{table7}.

We take $\chi_{cJ}\to \phi\phi$ as an representative. By regarding
$\phi$ as a strangeonium,  we can directly use Eq.~(\ref{decay1}), only
with some trivial changes of input parameters.
 We take the constituent quark mass $m_s \approx
M_\phi/2 = 0.5$ GeV. The radial wave function at the origin of $\phi$,
$|R_{\phi}(0)|^2$, can be extracted analogously from its measured
dielectron width of $0.19$ GeV$^3$.   Taking $M_{\chi_{c0}}=3.41475$
GeV,  and the strong coupling constant
 $\alpha_s=0.3$, and only considering  the
contribution of the leading-order QCD and QED contributions, we can obtain the decay width and
branching ration of $\chi_{cJ}\to \phi \phi$ in Table~\ref{table8}.

It is also interesting to consider the similar decay process $\chi_{cJ}
\to \omega \omega$,   Parallel to the preceding procedure, we also give
the results in Table~\ref{table8}.

From the Table~\ref{table8}, we can see our predictions of
Br$[\chi_{c0}\to \phi\phi]$ and Br$[\chi_{c2}\to \phi\phi]$ are $3.2\times 10^{-4}$ and $30\times 10^{-4}$ respectively. Our results are compatible with  the results obtained in~\cite{Braguta:2005gw} within ``$\phi_3$" model~\cite{kartvelishvili:1978} not only in decay width but also in branching ratio. Moreover, the branching ratio of  $\chi_{c0,2}\to \phi\phi$ is also compatible with BESIII~\cite{Ablikim:2011} measurement and PDG~\cite{pdg} in the order of magnitude.
For the process $\chi_{cJ}\to \omega \omega$,
We  can  learn from Table~\ref{table8} that our result of Br$[\chi_{c0}\to \omega\omega]$ is
close to the observation in BESIII~\cite{Ablikim:2011}, while for Br$[\chi_{c2}\to \omega\omega]$, our result is close to the PDG~\cite{pdg}. It is noting that both  of the observation in BESIII~\cite{Ablikim:2011} and the result in PDG~\cite{pdg}, the branching ratio of  $\chi_{c0}\to \omega\omega$ is large than the one of $\chi_{c2}\to \omega\omega$, while our prediction is opposite. This discrepancy may be caused by the model approximation.

\section{conclusion}\label{summarization}

The major task of this work is to calculate the first-order relativistic corrections and electromagnetic corrections of
$\chi_{bJ}\to J/\psi J/\psi$ in the context of NRQCD factorization.
We first introduced NRQCD factorization formula, which is particularly suitable for calculating the relativistic
corrections to quarkonium production and decay processes in the color-singlet channel. Then we compared the orthodox matching strategy with our matching method and emphasized that our approach avoids the complications that come from squared amplitude and phase-space integral. These two methods are equivalent thanks to the Gremm-Kapustin relation and we give our matching schema at amplitude level.

As a phenomenological application of our results, we calculate the decay width and branching ratio of $\chi_{bJ}\to J/\psi J/\psi$. The
conclusion is that the relativistic contributions of $\chi_{bJ}\to J/\psi
J/\psi$ are modest, about $13.8\%$ for $\chi_{b0}\to J/\psi
J/\psi$, and about $12.8\%$ for $\chi_{b2}\to J/\psi J/\psi$. The
QED contributions are very small, only about $0.2\%$ for $\chi_{b0}\to
J/\psi J/\psi$ and $\chi_{b2}\to J/\psi J/\psi$. The more interesting branching ratios are predicted to be of order $10^{-5}$
for $\chi_{b0}\to J/\psi J/\psi$, $\chi_{b2}\to J/\psi J/\psi$, but
$10^{-11}$ for $\chi_{b1}\to J/\psi J/\psi$, since there is only
electromagnetism contribution in this channel. With the result, we
predict it is possible to observe these processes in LHC.

Finally, as an exploratory work,  we also have a quick
order-of-magnitude estimate of the decay width and branching ratio of
$\chi_{cJ}\to \omega\omega,\phi\phi$ by our formula in
leading-order QCD and QED contribution working with the constituent quark model. The predict of Br$[\chi_{c0,2}\to \phi\phi]$
is consistent with the experimental result in the order of magnitude.

\acknowledgments
We thank Yu Jia for proposing this research project, for many useful discussions, and for reading the manuscript. We thank Jia Xu for checking parts of the calculations. We thank Changzheng Yuan for bringing Ref.~\cite{Ablikim:2011} to our attention. The research was partially supported by the National Natural Science Foundation of China under Grant No.~10875130 and 10935012, and by China Postdoctoral Science Foundation.



\end{document}